\documentstyle[aps,preprint,prb]{revtex}

\def\ia{\AA$^{-1}$}
\def\u2{$\langle u^{2} \rangle$}

\begin{document}

\title{Possible charge inhomogeneities in the CuO$_2$ planes of 
YBa$_{\bf 2}$Cu$_{\bf 3}$O$_{\bf 6+x}$ (x=0.25, 0.45, 0.65, 0.94) from pulsed 
neutron diffraction}

\author { M. Gutmann and S. J. L. Billinge } 
\address{ Department of Physics and Astronomy and Center for
          Fundamental Materials Research,\\
          Michigan State University, East Lansing, Michigan 48824-1116.}
\author { E. L. Brosha and G. H. Kwei}
\address{Los Alamos National Laboratory, Los Alamos, New Mexico 87545.} 

\date{\today}

\maketitle


\begin{abstract}
The atomic pair distribution functions (PDF) of four powder samples of 
YBa$_2$Cu$_3$O$_{6+x}$ (x=0.25, 0.45, 0.65, 0.94) at 15 K have been 
measured  by means of pulsed neutron diffraction. The PDF is modelled
using a full-profile fitting approach to yield structural parameters.   
In contrast to earlier XAFS work we find no evidence of a split 
apical oxygen site.  However, a slightly improved fit over the 
average crystallographic model results when the planar Cu(2) site is split
along the $z$-direction. 
This is interpreted in terms of charge inhomogeneities in the
CuO$_2$ planes.
\end{abstract}



\section{Introduction}

The observation of charge stripes\cite{tranq;n95} in 
La$_{2-x-y}$Nd$_{y}$Sr$_{x}$CuO$_{4}$ raises the interesting possibility
that inhomogeneous charge distributions in general, and stripes in particular,
are a generic phenomenon of high-temperature superconductors.  Charge
stripes give rise to local structural distortions which can be evident
using local structural probes.  For example,
both atomic pair distribution function
(PDF) analysis on neutron powder diffraction 
data\cite{bozin;prl99;unpub,bozin;prb99}
and extended x-ray absorption fine structure (XAFS)\cite{bianc;prl96}
indicate that in the La$_{2-x}$Sr$_{x}$CuO$_{4}$ system local 
structural distortions exist which are consistent with such charge 
inhomogeneities.
It is clearly important to establish their presence more widely in the 
high-temperature superconductors.

A long-standing controversy exists between the diffraction and XAFS  
communities concerning the
existence of a double well potential for the apical, O(4), ion in
YBa$_2$Cu$_3$O$_{6+x}$ (123).  This was first reported from XAFS 
data\cite{conra;s89,mustr;prl90} as a 
double well with the minima of the wells separated
along the $c$-direction by 0.13~\AA .  It was also reported that the
well structure becomes modified close to 
T$_c$.\cite{conra;s89,mustr;prl90,mustr;prb92}  
The controversy arose 
because this result seemed to contradict single crystal\cite{sulli;prb93,schwe;prb94} 
and powder 
diffraction\cite{franc;ssc88,willi;prb88,kwei;pc90,kwei;pc91} 
data which gave no
evidence of enlarged thermal factors for O(4) along the $z$ direction
as would be expected if such a double well existed.  In addition, the XAFS 
result predicts an anomalously short
Cu(1)-O(4) bond (Cu(1) is the chain copper)\cite{egami;pms94} which seems
to be questionable on chemical grounds.  Nonetheless, subsequent XAFS studies
have consistently reproduced the main result: that the Cu(1)-O(4) and
Cu(2)-O(4) pair
distributions from the data are best modeled as each having
two equally populated components separated
by $\sim 0.1$~\AA .\cite{stern;pc93,booth;prb96,tyson;pc97}  There appears to 
be no correlation between superconducting properties and the observation of the
split position.\cite{stern;pc93,booth;prb96,tyson;pc97}  
Also, the split Cu-O(4) correlations are not present in all
samples.\cite{stern;pc93,booth;prb96}  
The importance of this structural feature to the superconductivity
is clearly doubtful; however, a solution to this controversy may
elucidate important information about the properties of these 
materials, especially in light of the significant evidence that lattice
effects are important in the 
superconductors.\cite{egami;pms94,egami;b;pphtsv96}

We have taken a different approach to study this problem.  We have
made an atomic pair distribution function (PDF) analysis of neutron powder
diffraction data.  The PDF technique is a diffraction technique which,
nonetheless, reveals local atomic structure directly.\cite{egami;b;lsfd98}  
In this sense it bridges the diffraction and local structure domains.
We would expect the PDF to reflect the atomic pair distributions 
observed with XAFS whereas a conventional Rietveld analysis of the same
data set should recover the crystallographic result.

In the PDF technique
the total scattering data are measured including both Bragg and diffuse
scattering.  These data are Fourier transformed into real space yielding
the PDF directly.  There are two main advantages of this approach over XAFS.
First, the data reduction to obtain the PDF is straightforward and deductive
and results in a virtually undistorted, high-resolution, PDF. This can
also be recovered from XAFS data, but only by careful fitting 
procedures.\cite{booth;prb96,zabin;prb95}
Second, the PDF is
obtained over a wide range of atomic separation, $r$.  This allows data
modeling to be carried out over an extended range of the PDF which puts
more constraints on possible data interpretations and makes structural
solutions more (though not completely) unique.  The disadvantage
with respect to XAFS  of the
present study, is that the total PDF is measured
rather than a chemical specific PDF which has fewer atom-pairs
contributing to the observed PDF.  Also, most  XAFS studies
relating to this question were made on oriented samples with a polarized
beam which further reduces the number of correlations in the resulting
PDF.\cite{conra;s89,mustr;prl90,mustr;prb92,%
stern;pc93,booth;prb96,tyson;pc97}  Whilst this kind of analysis is, in
principle, possible using diffraction\cite{egami;b;lsfd98} it has not been
done.

We have measured PDFs from four samples of YBa$_2$Cu$_3$O$_{6+x}$ 
with x=0.25, 0.45, 0.65, 0.94 at 15K.  These were analyzed using
PDFFIT,\cite{proff;jac99;unpub}
a full-profile fitting 
program analogous to Rietveld refinement but which fits the PDF and
therefore yields local structural information.\cite{billi;b;pom98}
The resulting structural
parameters are in good agreement with the average crystal structure; in
particular we refine a thermal factor on the O(4) site which is
not unphysically large.  We have attempted to refine split positions
on O(4) without success.  However, we do refine a split position along $z$ on
the in-pane copper site which results in a small improvement in
agreement.   We interpret
this observation in terms of an inhomogeneous charge distribution in the
CuO$_2$ plane consistent with the presence of localized charges; for
example, as would be expected in the presence of charge stripes.


\section{Experimental}

The YBa$_2$Cu$_3$O$_{6+x}$ samples were prepared using standard solid 
state reaction methods. Stochiometric quantities
of CuO, BaCO$_3$, and Y$_2$O$_3$ were ground in
an Al$_2$O$_3$ mortar and pestle under acetone until well mixed. The sample was 
air-dried and the powder contents
were loaded into a 3/4" diameter steel die and uni-axially pressed at 1000 
lbs. The pellet sample was removed and
placed into an alumina boat. The sample was placed into a preheated 
(800~$^\circ$C)-tube furnace under 1 atm of
flowing oxygen. The furnace tube was sealed and the furnace temperature was 
immediately raised to 960~$^\circ$C. The
sample was fired for 72 hours and quenched. After cooling to room 
temperature, the sample was removed,
ground under acetone, and repressed. This firing cycle was repeated until the 
powder x-ray diffraction traces
showed no evidence of the presence of second phases. The typical reaction 
time was on the order of 1 week. Fully
oxidized YBCO ($x=0.94$)
was prepared by heating a portion
of the YBCO sample to 450~$^\circ$C in 1 atm PO$_2$. The sample was cooled to room 
temperature at a rate of 2~$^\circ$C/min
and at 1 atm PO$_2$. 

The oxygen non-stoichiometric YBCO samples were prepared using data from 
Kishio {\it et al.}\cite{kishi;mrssp89} Portions of the first 
YBCO sample were divided and placed into 
separate alumina boats. The annealing
temperature and PO$_2$ were determined from Kishio~{\it  et al}. as set by the desired 
oxygen content for each sample.
An Ametek oxygen analyzer was used to set the oxygen content of the annealing 
gas mixture as determined from
the furnace exhaust while the furnace was at room temperature. The oxygen 
concentration was maintained by
diluting a 20\% O$_2$/Ar balance gas mixture with Ar. Each sample was annealed at 
the selected temperature until
the oxygen content of the exhaust gas returned the concentration that was 
intially set by the mixing manifold.
Typically, upon heating, the samples would lose oxygen and this would cause a 
spike in the measured oxygen
concentration. After equilibration, the boat was quenched to room temperature 
under the controlled PO$_2$
atmosphere with the aid of a Pt wire that passed through a septum cap to the 
furnace tube exterior. Thus the
atmosphere inside the furnace was maintained during the entire anneal and 
subsequent quench to room
temperature. 

The oxygen stoichiometry of each sample was determined by reduction of the 
sample in 6\% H$_2$/Ar forming gas.
The TGA scans were made at 5~$^\circ$C/min to a maximum temperature of 1100~$^\circ$C. The 
gas flow rate was 80 ml/min. 
A high resolution Siemens D5000 powder x-ray diffractometer using Cu Ka 
radiation and an incident beam
monochrometer was used for XRD. A Perkin/Elmer TGA 7 was used for 
thermogravimetry. 

Neutron powder diffraction data were collected on the High Intensity
Powder Diffractometer (HIPD)
at the Manuel Lujan Neutron Scattering Center 
(MLNSC) at Los Alamos National Laboratory for the 
$x=0.65$ and 0.94 samples and on the Glass, Liquids and Amorphous 
diffractometer (GLAD) at
the Intense Pulsed Neutron Source (IPNS) at Argonne National Laboratory
for the $x=0.25$ and 0.45 samples.
The samples of about 10 g were sealed in a cylindrical 
vanadium tube with helium exchange gas. Data were collected 
at various temperatures between 15~K and room temperature 
in a closed cycle helium refrigerator. 
Additional data sets were collected
to account for the scattering from the sample environment and the
empty can. A vanadium rod was measured to account for the flux
distribution at the sample position.
The data are corrected for detector deadtime and efficiency, 
background, absorption, multiple scattering, inelasticity effects and
normalized with respect to the incident flux and the 
total sample scattering cross-section to yield the total scattering
structure function, $S(Q)$.  This quantity is 
Fourier transformed
according to 
\begin{equation}
  G(r)= \frac{2}{\pi }\int_{0}^{\infty} Q [S(Q)-1] \sin (Qr)\> dQ.
\end{equation}
Data collection and analysis procedures have been described
elsewhere.\cite{billi;prb93} Random errors in the data from statistical
counting fluctuations are estimated by propagating the errors from the
raw data using standard error propagation.\cite{prince;book} The error
propagation process has been described in detail elsewhere 
.\cite{billi;PhD,toby;aca90} The reduced structure factor $F(Q)=Q[S(Q)-1]$ 
from a typical data-set is shown in Fig.~\ref{fig;data}(a) 
%
%
The resulting PDF is shown in Fig.~\ref{fig;data}(b). 

The PDF is a real-space representation of the local structure in the 
form of pair distances.  Modeling of PDF was carried out using 
the PDFFIT program\cite{proff;jac99;unpub} 
to perform a least-squares full-profile fit.
The 
structural inputs for the program are atomic positions, occupancies 
and anisotropic thermal factors directly analogous to Rietveld refinements,
allowing direct comparison of data analyzed in real and reciprocal
space.  Estimated standard deviations on the refined values are obtained
from the variance-covariance matrix in the usual way.\cite{prince;book} 
Again, we stress that the PDF is being fit and the local
structure is obtained from the PDF refinement.  This is because the PDF is 
obtained from both  Bragg and diffuse scattering whereas the Rietveld 
fits include only Bragg scattering. We have also carried out Rietveld
refinement of our data in reciprocal-space using the 
program GSAS.\cite{larso;unpub87} and these are compared with the
PDF fits.
Local distortions away from the average structure are incorporated
into the PDF modeling by reducing the symmetry or increasing the 
size of the unit cell used in the model.  


\section{Results}

\subsection{Comparison with the crystallographic structure}

We would like to know if refinements of the PDF reproduce the average
crystallographic structure. We compare our refinements with the results 
of the single-crystal x-ray diffraction study of Schweiss {\it et al.}.  
The results are shown in Table~\ref{tab;usvsthem}.
The data are from the the $x=0.94$ sample
taken at 90~K and the constraints on the atomic positional and anisotropic
thermal factors were made to mirror those of the single crystal study.
We took two separate data-sets at 90~K on cooling and warming.  The
results from each data-set reproduce very well so only the cooling cycle
data-set refinements are reproduced in the table.  The agreement with the
single crystal study is clearly very good. 

We have also compared our PDF fits with Rietveld refinements of our own 
data using the
GSAS Rietveld package.\cite{larso;unpub87}  Again, the agreement is
good although the Rietveld refinement produced some unphysical
thermal factors.
In the case of the PDF fits the data are used over a
wider range of $Q$ than the Rietveld fits: $Q_{max}=25$~\ia\ for the PDFs 
and $Q_{max}=15.7$~\ia ($d=0.4$~\AA) for the Rietveld refinements.  
The qualitative results of the average structure are reproduced in
the local structure; for example, U$_{33}$ on the apical oxygen site
is small (0.0046(2))\cite{123note} in both the Rietveld and PDFFIT refinements.  
The good agreement between the PDF and crystallographic fits
suggest that the difference between the XAFS and crystallographic
results cannot be explained simply due to the different
length-scale of the two measurements.

\subsection{Comparison with the XAFS results}

In Fig.~\ref{fig;XAFSmodel}(a)
%
%
 we compare two PDF models which simulate the
crystallographic and XAFS models.  The PDF calculated using
the average crystal structure is shown as a solid line.  The dashed line
shows the PDF calculated when the O(4) site is split by 0.13~\AA\  along $z$
and each site is occupied 50\% as suggested by the XAFS results.  All other
parameters in the models were kept the same.  The difference is shown below.
The presence of such a split position clearly has a small but
significant effect 
on the PDF throughout the $r$-range.  
The magnitude of the signal expected from a split O(4) site
is small in the total-PDF; however by fitting over a range of $r$
it should be possible to establish the existence of such a split from
the PDF if it exists.

In Fig.~\ref{fig;XAFSmodel}(b) we show the data from
$x=0.94$, $T=90$~K, as symbols
with the PDF from the converged crystallographic model 
plotted as a solid line.  This is the same as the solid line in 
Fig.~\ref{fig;XAFSmodel}(a).  The difference curve is shown below. 
If the split site exists in the data but not in the model,
which is constrained to have the crystallographic symmetry with no split
site, we would expect the two difference curves to be similar. 
The difference curves appear uncorrelated suggesting qualitatively
that the split O(4) site is not present in the data.

\subsection{Search for anharmonic atomic sites}

Using PDFFIT we searched for possible split positions in the structure.
First, motivated by the XAFS model, we tried refining a split position
for the O(4) ion along the $z$-direction.  Refinements were carried out
over the range $1\le r\le 15$~\AA\ and $1\le r\le 5$~\AA\ to check the
response of the intermediate and short-range structure.  An initial split
of 0.1~\AA\ was given to the model and $\Delta $, the magnitude of the split,
was allowed to vary.  The occupancy of each split site was set initially to
0.5 but allowed to vary with the constraint that $n_A+n_B = 1.0$ where
$n_A$ and $n_B$ are the occupancies of the $A$ and $B$ sites respectively
(see Fig.~\ref{fig;ourmodel} for details). 
In both cases $\Delta $ refined to a negligible value, again
questioning the existence of such a split site.

Based on the crystallographic results we notice that the chain oxygen,
O(1), has a large thermal factor along the $a$ direction perpendicular to
the chain.  Also Cu(1), the chain copper, has a relatively large thermal
factor perpendicular to the chain.  Also, interestingly, the in-plane
copper, Cu(2), has a similarly large thermal factor along $z$.  We searched
for evidence of split atomic sites on each of these atoms in the directions
indicated taking a similar approach to that used for the O(4).  Interestingly,
the only case where a split site refined to a finite value giving a lower
residual was in the case of Cu(2) split along $z$.  The results are summarized
in Tables~\ref{tab;unsplit} and ~\ref{tab;cusplit}.
The geometry of the split Cu(2) $z$ sites are shown in Fig.~\ref{fig;ourmodel}.
%
%
The two distinct copper sites are labeled A and B corresponding to sites 
which are displaced respectively towards and away from the apical oxygen.

For fits over the range to 15.27~\AA ,
refinements to all the data-sets were stable and convergent for
the split position on Cu(2) along $z$ 
(which was not the case for a split on O(4) for example);
however, the split sometimes refined to a negligibly small value and the
improvement in fit is always small and barely significant.  The results
are much more robust when the fit is confined to the range 
$1.5 < r <  5.2$.  In this case significant improvements in agreement factor
are produced when splits in the range $x=0.10(5)-0.28(8)$~\AA\ are refined.
Refined values from the GLAD data and the HIPD data are self consistent
but there is not quantitative agreement between the results from
these two instruments.  The GLAD data refine a  significantly larger
split.  This is consistent with the observation in Table~\ref{tab;unsplit}
that GLAD thermal factors are consistently larger than those refined from
HIPD data.  The origin of this is that the GLAD data have a lower real-space
resolution.  We correct in the modeling for the experimental resolution
coming from the finite $Q_{max}$ of the measurement; however, there is an
additional loss in real-space resolution which comes
from a $Q$-dependent asymmetric line broadening from time-of-flight
spectrometers\cite{toby;ac92} which is significant in the GLAD data
and is not, at present, corrected in the modeling.  Thus, we expect the
GLAD data to overestimate the split.  The most likely value of the split
is between 0.1 and 0.2~\AA  . The observation of disorder on the Cu(2) site
but not the apical oxygen site is consistent with another recent differential
PDF study on
optimally doped YBa$_2$Cu$_3$O$_{6+x}$.\cite{louca;prb99;unpub}

\section{Discussion}

First we discuss the inability to refine a split site on O(1) perpendicular
to the chain.  The thermal factors are enormous on this site in this
direction and there is certainly lateral disorder associated with the
chain oxygens.  It was therefore a surprise that a split position did not
improve the fit.  Presumably the reason is that the chain is buckled in 
such a way that the oxygen ions take up a variety of displaced positions
rather than two distinct displaced positions.  This implies that the 
wavelength of the buckling is longer than two unit cells.  
We cannot determine if this buckling is
static or dynamic.  A similar argument can be made for the Cu(1) displacements
in the $a$ direction, though these are somewhat smaller than those observed
on the chain oxygen ions.

We turn now to the planar copper ions.  In this case a finite split is refined
with an improvement in residual.  Based on preceding arguments, this
suggests that the distribution of Cu(2) is bimodal (Fig.~\ref{fig;ourmodel}).  
This might
be expected if every copper site is not in the same charge-state as would
happen in the presence of polarons or local charge-stripes in which
case some sites would be Cu$^{2+}$ and others Cu$^{3+}$, for example.
Cu(2) sits at the base of a pyramidal cap of oxygen ions and does not lie
on a center of symmetry; this is in contrast to the copper in single layer
materials where it is octahedrally coordinated.  In the present case,
in response to increasing its charge state, the copper might be expected
to change its position in such a way as to move towards the negative 
oxygen ions:  Cu$^{3+}$ is expected to move into the oxygen pyramid somewhat.
This is seen in the average structure as a function of oxygen concentration.
The average Cu(2)-O(4) bond length changes from 2.4421 \AA\  to 2.2708 \AA\  
on going  from $x=0.1$ to $x=0.94$. This effect is also seen
in XAFS.\cite{rohle;pc97} 
The difference in these two bond 
lengths
is 0.1713~\AA\ which is a similar magnitude to the observed split, $\Delta$.
If there are charge inhomogeneities in 
the CuO$_2$ plane copper ions with different Cu(2)-O(4) bond-lengths
will coexist.  In the absence of long-range
order this will only be evident directly in the local structure, though
an elongated thermal factor will be apparent crystallographically.
Thus, the PDF results are consistent with crystallographic observations
of an enlarged Cu(2) U$_{33}$,\cite{schwe;prb94,kwei;pc90,kwei;pc91}
(at least on underdoped samples\cite{schwe;prb94}).
Recently, yttrium XAFS results also point towards a Cu(2) site split
along the $z$-direction.\cite{rohle;pb99;unpub}  In this case the split
is smaller ($0.05$~\AA ) and is correlated with displacements of O(2) and 
O(3) and probably the whole pyramid of O(2,3,4).  However, the
disagreement with this study
in the size of the split is probably not significant given the
uncertaintly of both measurements.  It would be interesting to
model correlated atom displacements; however, we plan to collect
data with better statistics or differential PDF data before attempting this.
Correlated displacements of Cu(2) in directions out of the 
plane are also consistent
with the presence of diffuse scattering in electron 
diffraction.\cite{ether;pma96}
We also note that there is independent evidence for disorder on copper
sites which may be correlated with the charge-state of the copper. 
Ion-channeling experiments\cite{sharm;prl96} 
found an anomaly in the Y-Ba-Cu signal in superconducting 123 around
T$_c$ but not in the Y-Ba signal of the same sample.

The motivation for this study was to see whether doping induces a 
double-well potential at the apical oxygen site as suggested from earlier 
XAFS work. Our results clearly rule out a local split site for 
O(4) along $z$.  However, we 
note that our findings are not necessarily in contradiction with the XAFS 
results. This can be understood as follows:
In an XAFS experiment the neighbourhood of a particular atomic species 
is probed. The single scattering paths correspond to pair distance 
distributions similarly as observed from PDF. However, one of the 
atoms in the pair is the photoabsorber. If a distance appears split 
it is therefore not unique which of the two atoms in the pair sits 
on a split site. Thus, in the present case the beat observed in the
Cu XAFS\cite{mustr;prl90} may possibly be explained by a split position
on Cu(2).  In principle this could be resolved by investigating 
the pair correlations between other atoms too. In practice this is 
difficult since typically in the XAFS data analysis multiple scattering 
paths involving triple- and higher scattering paths contribute 
significantly beyond the first two nearest-neighbour peaks. The multiple 
scattering paths are generally difficult to take into account and their 
number grows rapidly with increasing distance from the photoabsorber. 
We also note that the XAFS signal is affected by the valence of the
photoabsorber ion.  We are not aware of an analysis of the XAFS data
which takes into account the possibility that some Cu sites are 2+ and others
3+. This might also help to reconcile the XAFS and diffraction work.

It is interesting that refinement of the Cu(2) split position is more
robust when a narrower range of $r$ is fit.  This suggests that even by
15~\AA\ the structure resembles the average structure.  This suggests
that any charge inhomogeneities are atomic scale and there is no
evidence of even nanometer-scale charge phase separation.  At optimal
doping we would expect $\sim 15\% $ of Cu sites to contain 
holes.\cite{schne;prl92} In the stripe 
model\cite{tranq;n95,bozin;prb99} the distance of closest approach
of these holes would be $\sim 5.4$~\AA\ and the separation of
stripes would be $\sim 16$~\AA\cite{bozin;prb99} .  
Atomic pair correlations originating in
one stripe and terminating in an adjacent one will more resemble the
average structure than the local configuration either in a charge stripe
or between stripes.  Our PDF results support charge inhomogeneities on
this atomic length scale rather than some kind of 
longer length-scale phase separation.

Local structural inhomogeneities in La$_{2-x}$Sr$_x$CuO$_4$\cite{bozin;prl99;unpub}
disappear in the overdoped state.  This coincides with
 the point where the pseudogap transition, $T_p$, 
merges with $T_c$.  The suggestion is that below $T_p$ the charge dynamics 
are from fluctuating localized charges whereas above $T_p$ the carriers
are delocalized and electronically induced structural inhomogeneities go 
away.\cite{uemur;n93}  We note that $T_p$ falls rapidly to $T_c$ between 
$x=0.95$ and $x=1.0$.\cite{demsa;el99}  The optimum doping level is sometimes
inferred by the value of $T_c$ only and the oxygen content is not well 
characterized. However, $T_c$ versus doping is quite flat around optimum 
doping. A sample with a true doping level which falls slightly below optimum 
doping would be in the underdoped regime and therefore exhibit structural 
distortions. A slightly overdoped sample on the contrary would not exhibit
structural distortions. This might explain the fact that
for samples around optimal doping there is some disagreement about the 
existence or otherwise of local structural 
distortions.\cite{stern;pc93,booth;prb96}

The presence of atomic scale charge inhomogeneities 
in the CuO$_2$ planes are presently 
emerging as a common feature of high-T$_c$ cuprates. Inhomogeneous
local bucklings of the 
Cu-O bond have been observed in La$_{2-x}$Sr$_x$CuO$_4$\cite{bozin;prl99;unpub,bozin;prb99} and 
Nd$_{2-x}$Ce$_x$CuO$_4$.\cite{billi;prb93,ignat;jsr99} 
Recently stripes have been observed in 
hole-doped 214 cuprates.\cite{tranq;n95} Our results suggest that such inhomogeneities 
are also present in YBa$_2$Cu$_3$O$_{6+x}$. 
In the present analysis we address only the question 
whether sites are split or not and not the
possibility of short-range ordered patterns of atomic displacements
in the chains and planes.\cite{rohle;pb99;unpub,ether;pma96} 
The existence of stripes cannot be invoked directly from our analysis.
However, they can be interpreted as being due to the presence of atomic
scale charge
inhomogeneities consistent with the presence of 
a dynamic local stripe phase.


\section{Conclusions}

From our pair distribution analysis on YBa$_2$Cu$_3$O$_{6+x}$ (x=0.25, 0.45, 
0.65, 0.94) we find no evidence for a split apical oxygen site. A finite split
of the order 0.1~\AA\  is obtained for the in-plane Cu(2) along $z$ 
resulted in an
improvement of the fit over the average crystallographic model. 
The origin of such a split
can be explained assuming the presence of both Cu$^{2+}$ and Cu$^{3+}$
in the CuO$_2$ planes as might be expected in the presence of charge
stripes or polarons.


\acknowledgements
We would like to thank V. Petkov, Th. Proffen and T. A. Tyson for 
invaluable
discussions and J. Johnson for her help with GLAD data collection.  
This work was supported by the NSF through grant DMR-9700966. 
MG acknowledges support from the Swiss National Science Foundation.  
The IPNS is funded by 
the U.S. Department of Energy under contract W-31-109-Eng-38 and 
the MLNSC under contract W-7405-ENG-36.


\bibliographystyle{/u24/billinge/bib/aip_simon}

\begin{thebibliography}{37}
\bibitem{123note}
This value is different from that shown in Table~\protect\ref{tab;usvsthem}
  (though they agree within the errors) because the Rietveld and single crystal
  refinements were carried out with slightly different constraints on thermal
  factors. Results of all Rietveld and PDFFIT refinements are available on
  request from the authors.
\end{thebibliography}

%
%
\begin{table}[!tb]
  \caption{Comparison between the structural parameters of YBa$_2$Cu$_3$O$_{6+x}$ 
at 90 K obtained from PDF refinements and single crystal 
data.\protect\cite{schwe;prb94} The 
PDF refinement was carried out over the range $1.5 < r < 15.27$~{\rm\AA}. 
The thermal parameters are in units of \AA$^2$. The numbers in parantheses 
 represent one standard deviation on the last digit. The space group used
 was $Pmmm$ with Y on (0.5,0.5,0.5), Ba on (0.5,0.5,$z$), Cu(1) on (0,0,0),
 Cu(2) on (0,0,$z$), O(1) on (0, 0.5, 0), O(2) on (0.5, 0, $z$), O(3) on (0, 0.5, $z$) and 
 O(4) on (0, 0, $z$).}
 \begin{tabular}{lll}
   Parameter                &PDF fits &Single crystal\\
 			   &	&Ref.~\protect\onlinecite{schwe;prb94}\\ 
   \hline
   a [{\AA}]                & 3.8269(6) & 3.8202 \\
   b [{\AA}]                & 3.8960(5) & 3.8858 \\
   c [{\AA}]                & 11.696(2) & 11.696 \\
   \hline
   Y U$_{11}$=U$_{22}$      & 0.0028(1) & 0.0032(4) \\
   U$_{33}$                 & 0.0040(2) & 0.0016(5) \\
   \hline
   Ba U$_{11}$=U$_{22}$     & 0.0018(1) & 0.0046(4) \\
    U$_{33}$                & 0.0047(2) & 0.0024(5) \\
    z                       & 0.1840(2) & 0.18367(12) \\
   \hline
   Cu(1) U$_{11}$=U$_{22}$  & 0.0038(2) & 0.0037(5) \\
    U$_{33}$                & 0.0025(2) & 0.0015(5)\\
   \hline
   Cu(2) U$_{11}$=U$_{22}$  & 0.0020(1) & 0.0020(3) \\
    U$_{33}$                & 0.0045(2) & 0.0029(4) \\ 
    z                       & 0.3548(2) & 0.35466(7) \\
   \hline
   O(1) U$_{11}$            & 0.0136(7) & 0.0116(15) \\
   U$_{22}$                 & 0.0051(4) & 0.0055(14) \\
   U$_{33}$                 & 0.0070(5) & 0.0047(11) \\
   Occupation               & 0.94      & 0.977(13) \\
   \hline
   O(2) U$_{11}$            & 0.0036(2) & 0.0037(4)\tablenote{We have adopted the naming scheme given in 
 Ref.~\protect\onlinecite{shake;b;csothtsco94} with O(2) at (${1\over 2}$,0,$z$),
 O(3) at (0,${1\over 2}$,$z$).  We 
 have converted the atomic positions
 from Ref.~\protect\onlinecite{schwe;prb94} to be consistent with this.} \\
   U$_{22}$                 & 0.0048(2) & 0.0046(4)\tablenotemark[1] \\
   U$_{33}$                 & 0.0052(2) & 0.0045(3) \\
   z                        & 0.3780(3) & 0.37818(7) \\
   \hline
   O(3) U$_{11}$            & 0.0048(2) & 0.0046(4)\tablenotemark[1]\\
   U$_{22}$                 & 0.0036(2) & 0.0037(4)\tablenotemark[1] \\
   U$_{33}$                 & 0.0052(2) & 0.0045(4) \\
   z                        & 0.3782(3) & 0.37818(7) \\
   \hline
   O(4) U$_{11}$=U$_{22}$   & 0.0043(2) & 0.0070(4) \\
   U$_{33}$                 & 0.0041(2) & 0.0041(5) \\
   z                        & 0.1599(2) & 0.15918(10) \\
   \end{tabular}
   \label{tab;usvsthem}
 \end{table}
\begin{table}[!tb]
  \caption{Structural data of YBa$_2$Cu$_3$O$_{6+x}$ from the 
           PDF refinements using the average crystallographic structure ($Pmmm$). All data were collected at 15~K.
           See Table~\protect\ref{tab;usvsthem} for atomic positions.}
  \label{tab;unsplit}
  \begin{tabular}{lllll}
                   & x=0.25    & x=0.45    &  x=0.65   & x=0.94    \\
  \hline
  Instrument       &GLAD       &GLAD       &HIPD       &HIPD       \\
  \hline
  a                & 3.8711(3) & 3.8556(7) & 3.8364(5) & 3.8256(7) \\
  b                & 3.8711(3) & 3.8884(7) & 3.8965(4) & 3.8944(6) \\
  c                & 11.807(2) & 11.749(2) & 11.724(1) & 11.693(2) \\
  \hline
   Cu(1)U$_{11}$   & 0.0052(2) & 0.0053(3) & 0.0028(2) & 0.0033(2) \\
   U$_{22}$        & 0.0052(2) & 0.0060(3) & 0.0045(3) & 0.0012(2) \\
   U$_{33}$        & 0.0012(2) & 0.0038(2) & 0.0025(2) & 0.0011(2) \\
  \hline
   Cu(2) U$_{11}$  & 0.0037(1) & 0.0025(2) &0.0015(1)  & 0.0015(1) \\
   U$_{22}$        & 0.0037(1) & 0.0049(2) & 0.0028(1) & 0.0021(1) \\
   U$_{33}$        & 0.0063(2) & 0.0059(2) & 0.0039(1) & 0.0032(2) \\
  z                & 0.3594(2) & 0.3581(2) & 0.3562(2) & 0.3564(2) \\
  \hline
  O(1) U$_{11}$    & 0.042(8)  & 0.030(2)  & 0.021(1)  & 0.0092(7) \\
  U$_{22}$         & 0.008(3)  &-0.0004(6) & 0.0027(4) & 0.0056(4) \\
  U$_{33}$         & 0.018(5)  & 0.0116(2) & 0.0012(2) & 0.0062(5) \\
 Occupation        &   0.25    &   0.45    &   0.65    &   0.94    \\
  \hline
  O(2) U$_{11}$    &0.0050(2)  & 0.0052(2) & 0.0035(1) & 0.0023(1) \\
  U$_{22}$         &0.0050(2)  & 0.0049(2) & 0.0046(1) & 0.0039(1) \\
  U$_{33}$         & 0.0074(2) & 0.0064(2) & 0.0050(1) & 0.0063(2) \\
  z                & 0.3780(2) & 0.3784(3) & 0.3791(3) & 0.3785(3) \\
\hline
 O(3) U$_{11}$     &0.0050(2)  & 0.0049(2) & 0.0046(1) & 0.0039(1) \\
 U$_{22}$          &0.0050(2)  & 0.0052(2) & 0.0035(1) & 0.0023(1) \\
  U$_{33}$         & 0.0074(2) & 0.0064(2) & 0.0050(1) & 0.0063(2) \\
  z(O3)            & 0.3780(2) & 0.3776(3) & 0.3777(3) & 0.3781(3) \\
  \hline
  O(4) U$_{11}$    & 0.0073(2) & 0.0067(3) & 0.0033(3) & 0.0020(2) \\
  U$_{22}$         & 0.0073(2) & 0.0079(3) & 0.0081(3) & 0.0042(3) \\
  U$_{33}$         & 0.0088(3) & 0.0073(3) & 0.0041(2) & 0.0055(2) \\
  z                & 0.1538(2) & 0.1554(2) & 0.1580(2) & 0.1601(2) \\
  \hline
  R$_{w}$         &    13.7   &   13.9    &   10.9    &    17.3   \\
  \end{tabular}
\end{table}
\begin{table}[!tb]
  \caption{Results of the PDF refinement including a split Cu(2) site
for various oxygen concentrations, $x$. 
The PDF refinements were carried out over the ranges $1.5 < r <  5.2$~{\AA} 
and $1.5 < r < 15.27$~{\AA}. $\Delta$ is the magnitude of the Cu(2) 
split along $z$, $n_A$ is the occupancy of the $A$-site in the splite
site model (see Fig.~\protect\ref{fig;ourmodel} for details), 
$R_{ws}$ and $R_{wu}$ are the weighted profile residuals
for the converged fits respectively with and without the Cu split position.
The numbers in parantheses represent one standard 
deviation on the last digit. }
  \label{tab;cusplit}
  \begin{tabular}{l|llll|llll}
&\multicolumn{4}{c|}{$1.5 < r <  5.2$~{\AA}}&\multicolumn{4}{c}{$1.5 < r <  15.27$~{\AA}}\\
  $x$    &${\Delta}$~(\AA )&   $n_A$ &  $R_{ws}$ &  $R_{wu}$ & ${\Delta}$ (\AA )&  $n_A$  &  $R_{ws}$&  $R_{wu}$  \\
  \hline
  0.25      & 0.28(8)  & 0.14(4) &  13.9&15.7  & 0.01(2)   &  0.50(3) &  13.3&13.7  \\
  0.45      & 0.20(6)  & 0.25(5) &  15.4 &16.9 & 0.17(6)   &  0.20(1)&  13.6&13.9  \\
  0.65      & 0.10(5)  & 0.85(8) &  11.9 &12.3 & 0.03(2)   &  0.10(4)&  10.7 &10.9 \\
  0.94      & 0.18(6)  & 0.87(8) &  21.4  &22.8& 0.15(2)   &  0.84(3)&  16.2 &17.3 \\
  \end{tabular}
\end{table}

\begin{figure}[!tb]
  \caption{(a) Reduced structure factor $F(Q)=Q[S(Q)-1]$ for 
YBa$_2$Cu$_3$O$_{6.45}$ measured
           at 15~K. (b) Reduced radial distribution function, $G(r)$,
from the data shown in (a).} 
  \label{fig;data}
\end{figure}

\begin{figure}[!tb]
  \caption{(a) PDF calculated using the average crystal
structure model (solid line) and the XAFS model of 
Ref.~\protect\onlinecite{mustr;prl90}(dashed line). 
(b) Open symbols: YBa$_2$Cu$_3$O$_{6.94}$ 90~K data, Solid line:
fit to the data using the average crystal structure model constrained in
the same way as the Rietveld refinements.  Differences are plotted below.} 
  \label{fig;XAFSmodel}
\end{figure}

\begin{figure}[!tb]
  \caption{Illustration of the split Cu(2) site. Two adjacent CuO$_5$ pyramids
are shown. The label for the A site is the same as used in 
Table~\ref{tab;cusplit}.  The size of the split is amplified for
clarity.}
  \label{fig;ourmodel}
\end{figure}

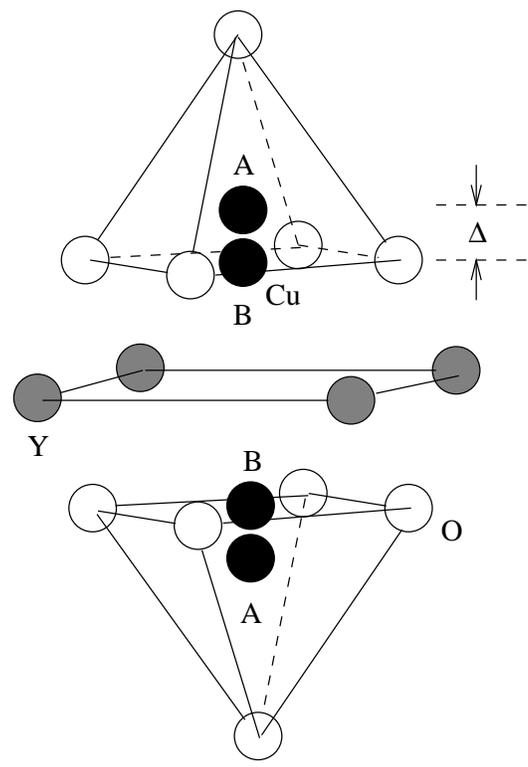
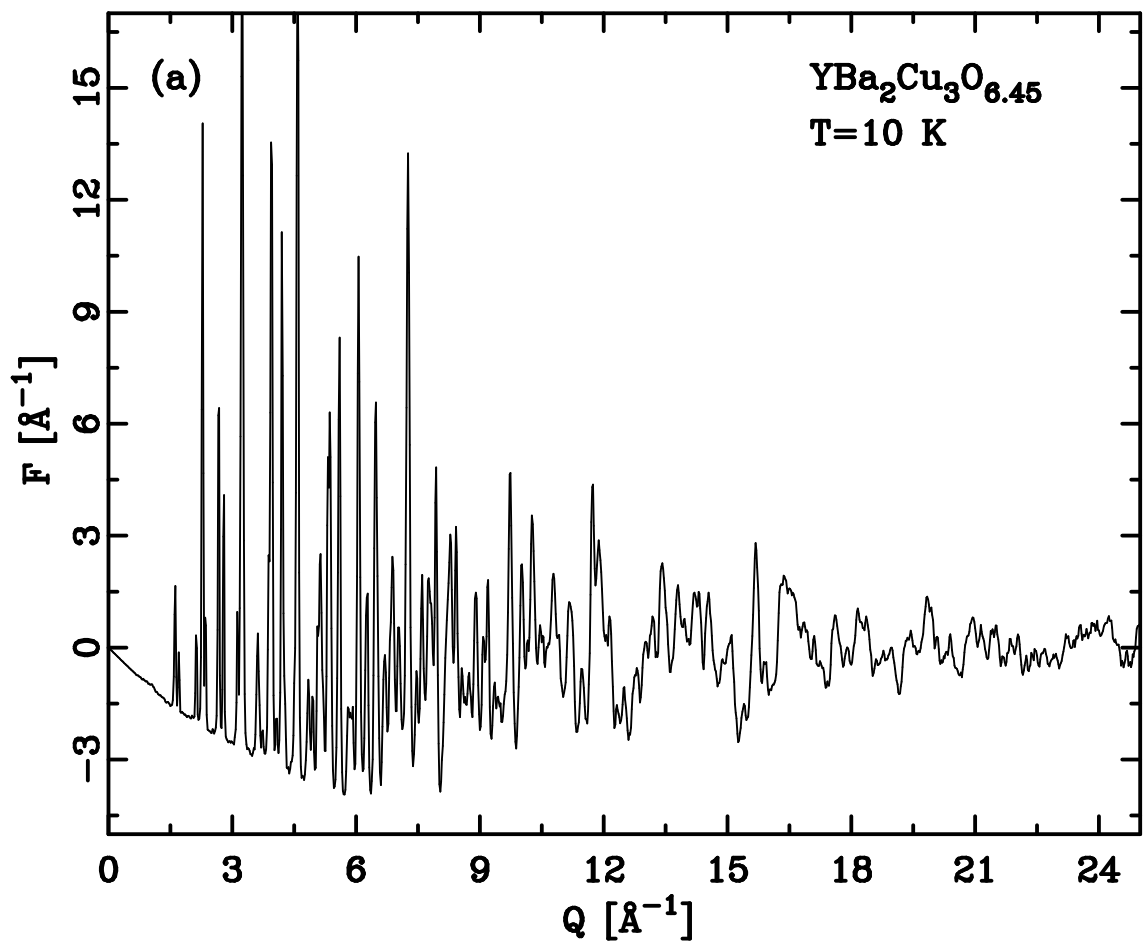
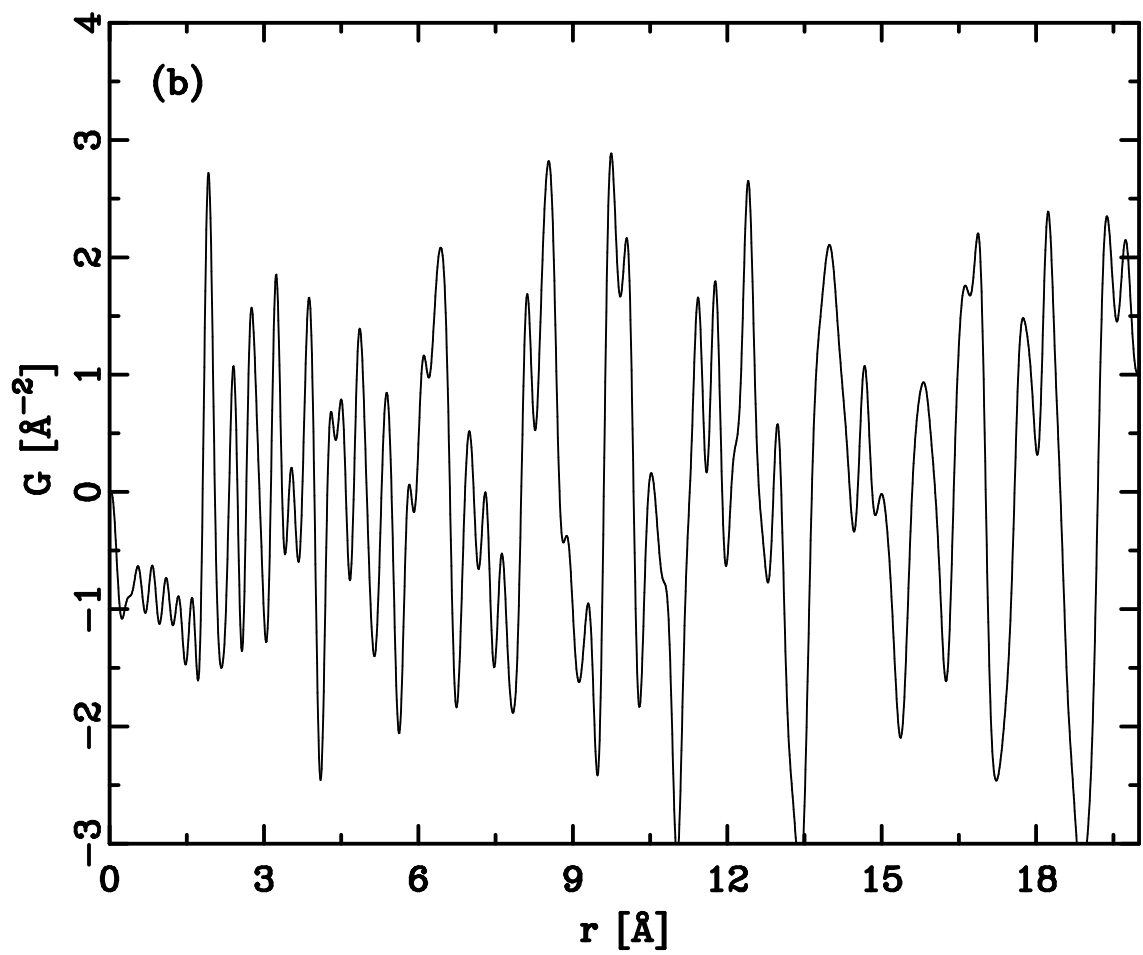
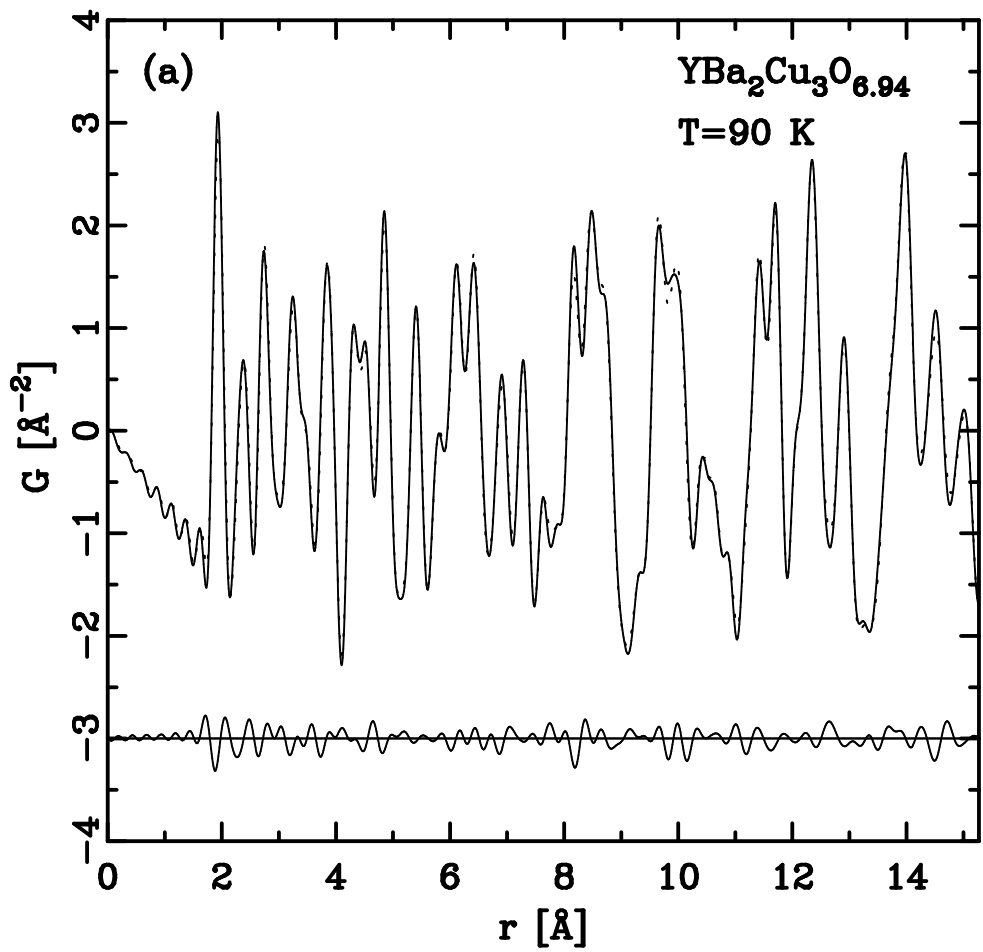
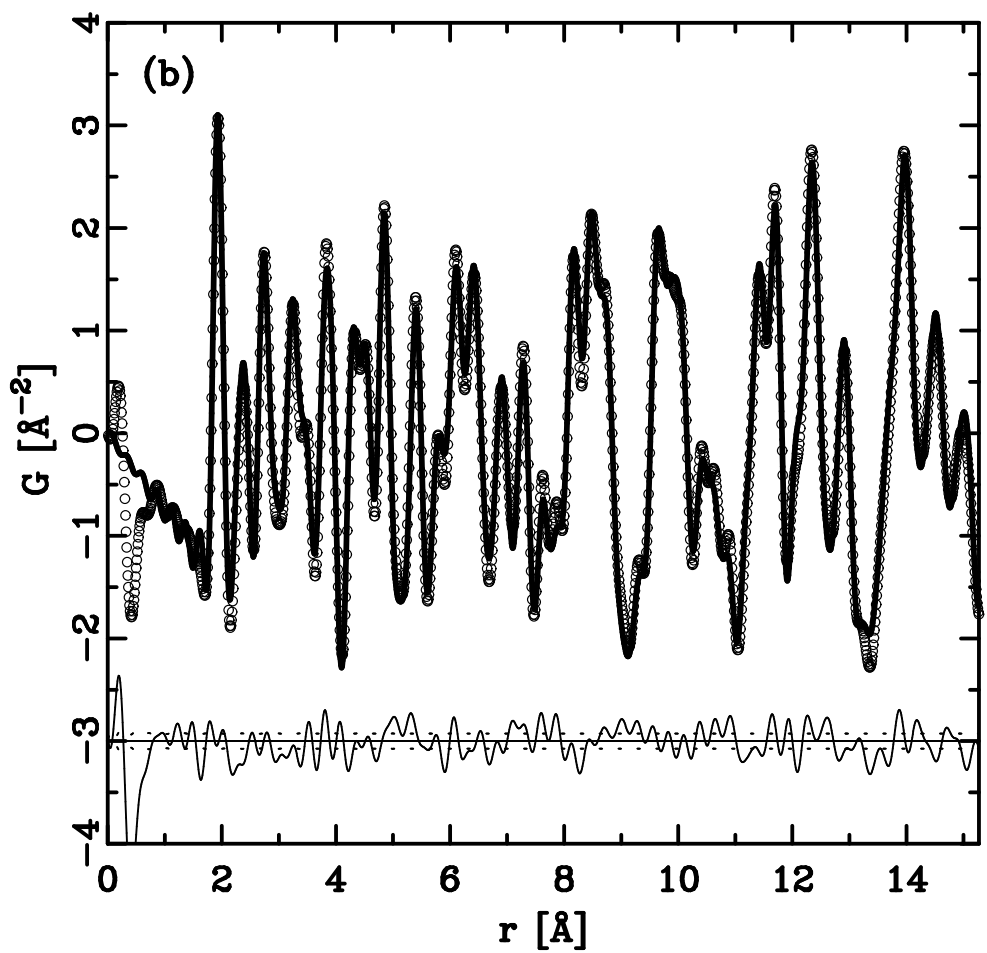
\end{document}